\documentclass[letterpaper,12pt]{article}

\usepackage{array}
\usepackage{jheppub}
\usepackage[all,pdf]{xy}

\def\ket#1{\langle #1 \rangle}

\DeclareMathOperator{\Conf}{Conf}
\DeclareMathOperator{\Gr}{Gr}

\title{
Symbol Alphabets from Plabic Graphs}

\author[1]{Jorge Mago,}\emailAdd{jorge\_mago@brown.edu}
\author[1]{Anders Schreiber,}\emailAdd{anders\_schreiber@brown.edu}
\author[1,2]{Marcus Spradlin}\emailAdd{marcus\_spradlin@brown.edu}
\author[1]{and Anastasia Volovich}\emailAdd{anastasia\_volovich@brown.edu}

\affiliation[1]{Department of Physics,
Brown University,
Providence, RI 02912, USA}

\affiliation[2]{Brown Theoretical Physics Center,
Brown University,
Providence, RI 02912, USA}

\abstract{Symbol alphabets of $n$-particle amplitudes in $\mathcal{N}=4$
super-Yang-Mills
theory are known to contain certain cluster variables of $\Gr(4,n)$
as well as certain algebraic functions of cluster variables.
In this paper we suggest an algorithm for computing these symbol alphabets
from plabic graphs by solving matrix equations of the form
$C \cdot Z = 0$ to associate functions on $\Gr(m,n)$ to parameterizations of
certain cells of $\Gr(k,n)$ indexed by plabic graphs.
For $m=4$ and $n=8$ we show that this association precisely reproduces the 18
algebraic symbol letters of the two-loop NMHV eight-particle amplitude from
four plabic graphs.
}

\begin{document}

\maketitle

\section{Introduction}

A central problem in studying
the scattering amplitudes
of planar $\mathcal{N}=4$ super-Yang-Mills (SYM) theory is
to understand their analytic structure.
Certain amplitudes are known
or expected to be expressible in terms of generalized
polylogarithm functions.  The branch points of any such amplitude
are encoded in its \emph{symbol alphabet}---a finite collection of
multiplicatively independent functions on kinematic space called
\emph{symbol letters}~\cite{Goncharov:2010jf}.
In~\cite{Golden:2013xva} it was observed that
for $n=6,7$,
the symbol alphabet of all (then-known) $n$-particle amplitudes
is the set of
cluster variables~\cite{FZ1,FZ2} of the $\Gr(4,n)$ Grassmannian cluster
algebra~\cite{Scott}.  The hypothesis that this remains true to arbitrary
loop order provides the bedrock underlying a bootstrap program
that has enabled the computation of these amplitudes to impressively
high loop order and remains supported by all available evidence (see~\cite{Caron-Huot:2020bkp} for a recent review).

For $n>7$ the $\Gr(4,n)$ cluster algebra has infinitely many
cluster variables~\cite{FZ2,Scott}.
While it has long been known that the symbol alphabets
of some $n>7$ amplitudes
(such as the two-loop MHV amplitudes~\cite{CaronHuot:2011ky})
are given by finite subsets
of cluster variables, there was no candidate guess for a ``theory'' to explain
why amplitudes would select the subsets that they do.
At the same time, it was expected~\cite{Prlina:2017azl,Prlina:2017tvx} that the symbol alphabets
of even MHV amplitudes for $n>7$ would generically require  letters that
are not cluster variables---specifically, that are algebraic functions of the
Pl\"ucker coordinates on $\Gr(4,n)$, of the type that appear in
the one-loop four-mass box function~\cite{Hodges1977,tHooft:1978jhc} (see Appendix~\ref{sec:roots}).
(Throughout this paper we use the adjective ``algebraic''
to specifically denote something that is algebraic but not rational.)

As often the case for amplitudes, guesses
and expectations are valuable but explicit computations are king.
Recently the two-loop eight-particle
NMHV amplitude in SYM theory was computed~\cite{Zhang:2019vnm},
and it was found to have a 198-letter
symbol alphabet that can be taken to consist of 180 cluster
variables on $\Gr(4,8)$ and an additional 18 algebraic letters that
involve square roots of four-mass box type.
(Evidence for the former was presented in~\cite{Prlina:2017tvx}
based on an analysis of the Landau equations; the latter
are consistent with the Landau analysis but less constrained by it.)
The result of~\cite{Zhang:2019vnm}
provided the first concrete new data on symbol alphabets
in SYM theory in over eight years.  We will refer to this
as ``the eight-particle alphabet'' in this paper since (turning again to hopeful speculation)
it may turn out to be the complete symbol
alphabet for all eight-particle amplitudes in SYM theory at all loop order.

A few recent papers have sought to explain or postdict
the eight-particle symbol alphabet and to clarify its
connection to the $\Gr(4,8)$ cluster algebra.
In~\cite{Arkani-Hamed:2019rds}
polytopal realizations of certain compactifications of
(the positive part of) the
configuration space
$\Conf_8(\mathbb{P}^3)$ of eight
particles in SYM theory were constructed.  These naturally select
certain
finite subsets of cluster variables, including those in the
eight-particle alphabet, and the square roots of four-mass box
type make a natural appearance as well.  At the same time, an equivalent but
dual description, involving certain fans associated to
the tropical totally positive Grassmannian~\cite{SW},
appeared simultaneously in~\cite{Drummond:2019cxm,Henke:2019hve}.
Moreover~\cite{Drummond:2019cxm} proposed a construction that
precisely computes the 18 algebraic letters of the eight-particle symbol alphabet
by (roughly speaking)
analyzing how the simplest candidate fan is embedded within
the (infinite) $\Gr(4,8)$ cluster fan.

In this paper we show that the algebraic letters
of the eight-particle symbol alphabet are precisely reproduced by an alternate
construction that only requires solving a set of
simple
polynomial equations associated to certain plabic (planar bi-colored) graphs.
Plabic graphs~\cite{Postnikov} are known
to encode important properties of SYM amplitudes~\cite{ArkaniHamed:2012nw},
in particular the singularities of integrands,
and our work
raises the possibility that symbol alphabets of SYM theory
could also be encoded in certain plabic graphs.
In Sec.~\ref{sec:mainidea} we introduce our construction with
a simple example and then complete the analysis for
all graphs relevant to
$\Gr(4,6)$ in Sec.~\ref{sec:sixparticle}.
In Sec.~\ref{sec:noncluster} we consider an example
where the construction yields non-cluster variables of $\Gr(3,6)$
and in Sec.~\ref{sec:rootsection} we apply it to graphs
that precisely reproduce the algebraic functions on $\Gr(4,8)$
that appear in the symbol of~\cite{Zhang:2019vnm}.

\section{A Motivational Example}
\label{sec:mainidea}

Motivated by~\cite{NimaTalk},
in this paper we consider
solutions to sets of equations of the form
\begin{align}
\label{eq:cdotz}
C \cdot Z = 0
\end{align}
which are familiar from the study of
several closely connected or essentially equivalent
amplitude-related objects
(leading singularities, Yangian
invariants, on-shell forms; see for example~\cite{ArkaniHamed:2009dn,Mason:2009qx,ArkaniHamed:2009vw,Drummond:2010uq,ArkaniHamed:2012nw}).

For the application to SYM theory that will be
the focus of this paper, $Z$ is the $n \times 4$ matrix of
momentum twistors describing the kinematics of an $n$-particle scattering
event, but it is often instructive to allow $Z$
to be $n \times m$ for general $m$.

The $k \times n$ matrix $C(f_0,\ldots,f_d)$ in~(\ref{eq:cdotz})
parameterizes a $d$-dimensional cell of the totally
non-negative Grassmannian $\Gr(k,n)_{\ge 0}$.
Specifically, we always take
it to be
the
boundary measurement of a (reduced, perfectly oriented) plabic graph
expressed in terms of the face weights $f_\alpha$
of the graph (see~\cite{Postnikov,ArkaniHamed:2012nw}).
One could equally well use edge weights, but using face weights
allows us to further restrict our attention to bipartite graphs
and to eliminate some redundancy; the only residual
redundancy of face weights is that they satisfy
$\prod_a f_\alpha = 1$ for each graph.

For an illustrative example, consider
\begin{equation}
\begin{aligned}
\includegraphics[width=1.5in]{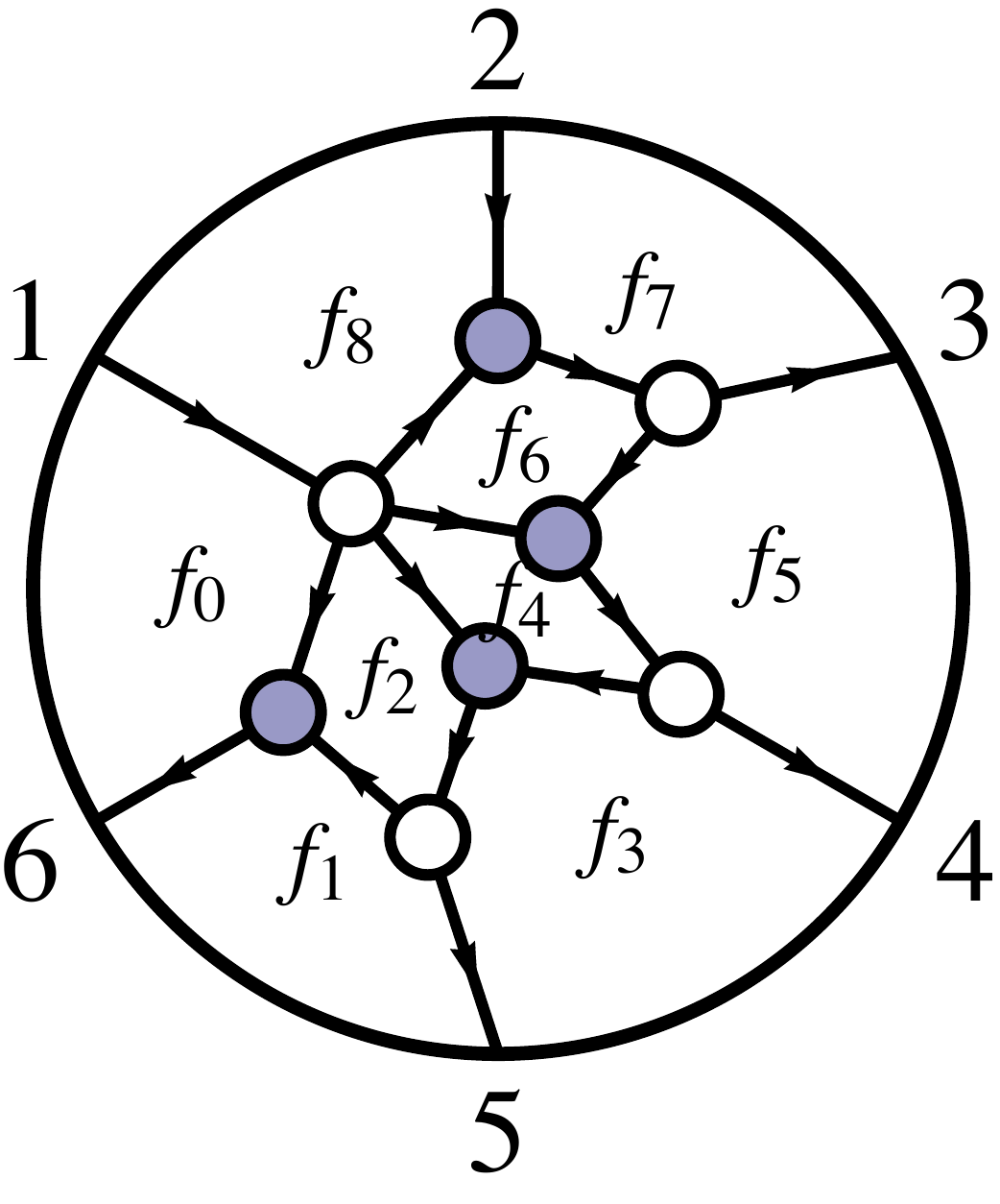}
\end{aligned}
\label{eq:examplegraph}
\end{equation}
which affords us the opportunity to review the construction of
the associated $C$-matrix from~\cite{Postnikov}.
The graph is \emph{perfectly oriented} because each black (white)
vertex has all incident arrows but one pointing in (out).
The graph has two sources $\{1,2\}$ and four sinks
$\{3,4,5,6\}$ and we begin by forming a $2 \times (2 + 4)$
matrix with the $2 \times 2$ identity matrix occupying the source
columns:
\begin{align}
C = \left(\begin{matrix}
1 & 0 & c_{13} & c_{14} & c_{15} & c_{16} \\
0 & 1 & c_{23} & c_{24} & c_{25} & c_{26}
\end{matrix}\right).
\label{eq:cmatrixform}
\end{align}
The remaining entries are given by
\begin{align}
c_{ij} = (-1)^s \sum_{p:i \mapsto j}
\prod_{\alpha \in \widehat{p}} f_\alpha
\end{align}
where $s$ is the number of sources strictly between $i$ and $j$,
the sum runs over all allowed paths $p$ from $i$ to $j$
(allowed paths must traverse each edge only in the direction of its arrow)
and the product runs over all faces $\alpha$ to the right of $p$,
denoted by $\widehat{p}$.
In this manner we find
\begin{equation}
\begin{aligned}
c_{13} &=-f_0 f_1 f_2 f_3 f_4 f_5 f_6\,, &
c_{23} &= f_0 f_1 f_2 f_3 f_4 f_5 f_6 f_8\,, \\
c_{14} &=-f_0 f_1 f_2 f_3 f_4 (1 + f_6)\,, &
c_{24} &= f_0 f_1 f_2 f_3 f_4 f_6 f_8\,, \\
c_{15} &=-f_0 f_1 f_2 (1 + f_4 + f_4 f_6)\,, &
c_{25} &= f_0 f_1 f_2 f_4 f_6 f_8\,, \\
c_{16} &=-f_0 (1 + f_2 + f_2 f_4 + f_2 f_4 f_6)\,, &
c_{26} &= f_0 f_2 f_4 f_6 f_8\,.
\end{aligned}
\end{equation}
Then for $m=4$, (\ref{eq:cdotz}) is a system of $2 \times 4 = 8$
equations for the eight independent face weights, which has the solution
\begin{equation}
\begin{aligned}
f_0 &= - \frac{\ket{1234}}{\ket{2346}}\,, &
f_1 &= - \frac{\ket{2346}}{\ket{2345}}\,, &
f_2 &= \frac{\ket{2345}\ket{1236}}{\ket{1234}\ket{2356}}\,, \\
f_3 &= - \frac{\ket{2356}}{\ket{2346}}\,, &
f_4 &= \frac{\ket{2346}\ket{1256}}{\ket{2456}\ket{1236}}\,, &
f_5 &=-  \frac{\ket{2456}}{\ket{2356}}\,, \\
f_6 &= \frac{\ket{2356}\ket{1456}}{\ket{3456}\ket{1256}}\,, &
f_7 &= - \frac{\ket{3456}}{\ket{2456}}\,, &
f_8 &= - \frac{\ket{2456}}{\ket{1456}}\,,
\end{aligned}
\label{eq:example1}
\end{equation}
where $\langle ijkl \rangle = \det(Z_i Z_j Z_k Z_l)$ are
Pl\"ucker coordinates on $\Gr(4,6)$.

We pause here to point out two features
evident from~(\ref{eq:example1}).
First, we see that on the solution
of~(\ref{eq:cdotz}) each face weight evaluates (up to sign)
to a product of powers of $\Gr(4,6)$ cluster variables, i.e.~to
a symbol letter of six-particle
amplitudes in SYM theory~\cite{Goncharov:2010jf}.
Moreover,
the cluster variables that appear ($\ket{2346}$, $\ket{2356}$, $\ket{2456}$,
and the six frozen variables)
constitute a single cluster of the $\Gr(4,6)$ algebra.

The fact that cluster variables of $\Gr(m,n)$ seem to arise,
at least in this example,
raises the
possibility that the symbol alphabets of amplitudes
in SYM theory might be given more generally by the face weights of
certain plabic graphs evaluated on solutions of $C \cdot Z = 0$.
A necessary condition for this to have a chance of working
is that the number of independent face weights should equal the
number of equations (both eight in the above example);
otherwise the equations would have no solutions or continuous families
of solutions.  For this reason we focus exclusively
on graphs for which~(\ref{eq:cdotz}) admits isolated
solutions for the face weights as functions of generic
$n \times m$ $Z$-matrices;
in particular this requires that $d=km$.
In such cases the number of isolated solutions to~(\ref{eq:cdotz})
is called the \emph{intersection number} of the graph.

The possible connection between plabic graphs and symbol
alphabets is especially tantalizing because
it manifestly has a chance to account for both issues
raised in the introduction:  (1) while the number of cluster
variables of $\Gr(4,n)$ is infinite for $n>7$, the number
of (reduced) plabic graphs is certainly finite for
any fixed $n$,
and (2) graphs with intersection number greater than 1
naturally provide candidate algebraic symbol letters.
Our showcase example of~(2) is presented in
Sec.~\ref{sec:rootsection}.

\section{Six-Particle Cluster Variables}
\label{sec:sixparticle}

The problem formulated in the previous section can be considered
for any $k$, $m$ and $n$.
In this section we thoroughly investigate
the first case of direct relevance to the amplitudes of SYM
theory: $m=4$ and $n=6$.  Although this case is special
for several reasons, it allows us to illustrate
some concepts and terminology that will be used in later sections.

\begin{figure}
\centering
\begin{tabular}
{
>{\centering\arraybackslash} m{0.3\textwidth}
>{\centering\arraybackslash} m{0.3\textwidth}
>{\centering\arraybackslash} m{0.3\textwidth}
}
\includegraphics[width=1.5in]{one.pdf}
&
\includegraphics[width=1.5in]{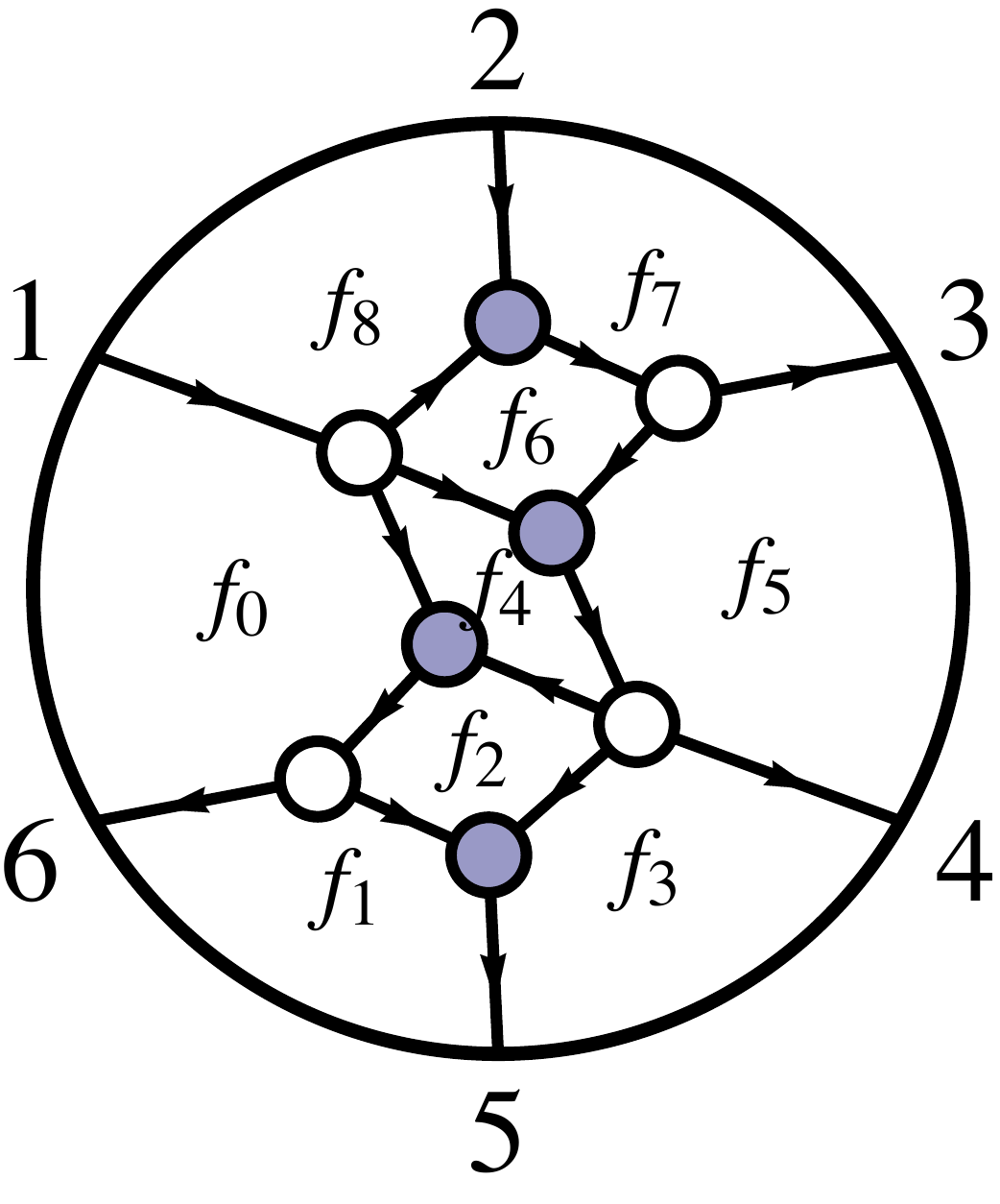}
&
\includegraphics[width=1.5in]{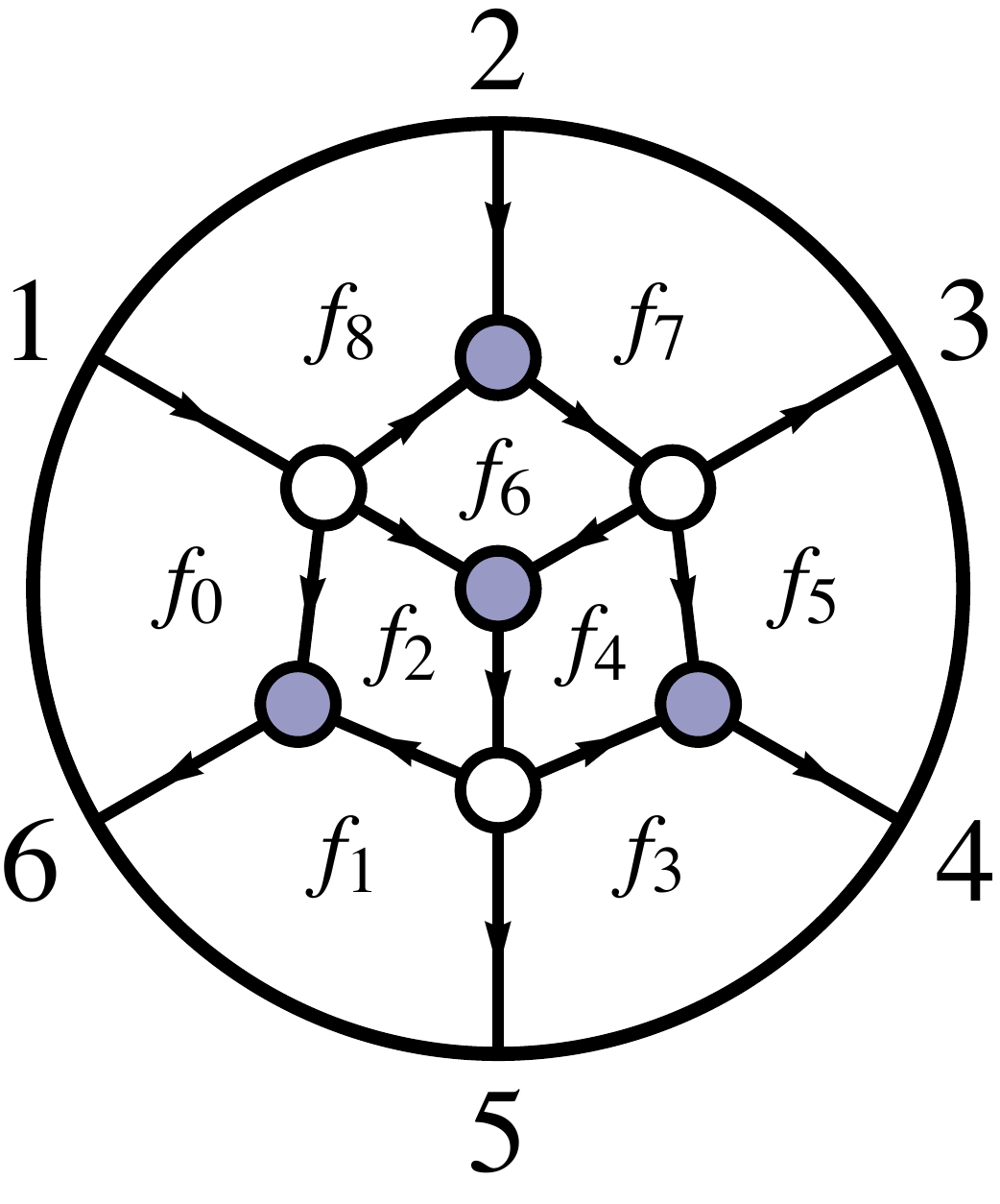}
\\
(a) & (b) & (c)
\\
\\
\\
\begin{xy} 0;<53pt,0pt>:<0pt,53pt>::
(-0.5,\halfrootthree) *+{\framebox[4ex]{$\scriptstyle{\langle 16\rangle}$}} = "a",
(0.5,\halfrootthree) *+{\framebox[4ex]{$\scriptstyle{\langle 12\rangle}$}} = "b",
(1,0) *+{\framebox[4ex]{$\scriptstyle{\langle 23\rangle}$}} = "c",
(0.5,-\halfrootthree) *+{\framebox[4ex]{$\scriptstyle{\langle 34\rangle}$}} = "d",
(-0.5,-\halfrootthree) *+{\framebox[4ex]{$\scriptstyle{\langle 45\rangle}$}} = "e",
(-1,0) *+{\framebox[4ex]{$\scriptstyle{\langle 56\rangle}$}} = "f",
(0.25,0.4) *+{\scriptstyle{\langle 26 \rangle}} = "g",
(0.1,-0.05) *+{\scriptstyle{\langle 36 \rangle}} = "h",
(-0.35,-0.4) *+{\scriptstyle{\langle 46 \rangle}} = "i",
"i", {\ar"h"},
"h", {\ar"g"},
"g", {\ar"a"},
"b", {\ar"g"},
"g", {\ar"c"},
"c", {\ar"h"},
"h", {\ar"d"},
"d", {\ar"i"},
"i", {\ar"e"},
"f", {\ar"i"},
"e", {\ar@{.>}"f"},
"a", {\ar@{.>}"b"}
\end{xy}
&
\begin{xy} 0;<53pt,0pt>:<0pt,53pt>::
(-0.5,\halfrootthree) *+{\framebox[4ex]{$\scriptstyle{\langle 16\rangle}$}} = "a",
(0.5,\halfrootthree) *+{\framebox[4ex]{$\scriptstyle{\langle 12\rangle}$}} = "b",
(1,0) *+{\framebox[4ex]{$\scriptstyle{\langle 23\rangle}$}} = "c",
(0.5,-\halfrootthree) *+{\framebox[4ex]{$\scriptstyle{\langle 34\rangle}$}} = "d",
(-0.5,-\halfrootthree) *+{\framebox[4ex]{$\scriptstyle{\langle 45\rangle}$}} = "e",
(-1,0) *+{\framebox[4ex]{$\scriptstyle{\langle 56\rangle}$}} = "f",
(0.2,0.4) *+{\scriptstyle{\langle 26 \rangle}} = "g",
(0,0) *+{\scriptstyle{\langle 36 \rangle}} = "h",
(-0.2,-0.4) *+{\scriptstyle{\langle 35 \rangle}} = "i",
"h", {\ar"i"},
"h", {\ar"g"},
"g", {\ar"a"},
"b", {\ar"g"},
"g", {\ar"c"},
"c", {\ar"h"},
"i", {\ar"d"},
"f", {\ar"h"},
"e", {\ar"i"},
"i", {\ar"f"},
"d", {\ar@{.>}"e"},
"a", {\ar@{.>}"b"}
\end{xy}
&
\begin{xy} 0;<53pt,0pt>:<0pt,53pt>::
(-0.5,\halfrootthree) *+{\framebox[4ex]{$\scriptstyle{\langle 16\rangle}$}} = "a",
(0.5,\halfrootthree) *+{\framebox[4ex]{$\scriptstyle{\langle 12\rangle}$}} = "b",
(1,0) *+{\framebox[4ex]{$\scriptstyle{\langle 23\rangle}$}} = "c",
(0.5,-\halfrootthree) *+{\framebox[4ex]{$\scriptstyle{\langle 34\rangle}$}} = "d",
(-0.5,-\halfrootthree) *+{\framebox[4ex]{$\scriptstyle{\langle 45\rangle}$}} = "e",
(-1,0) *+{\framebox[4ex]{$\scriptstyle{\langle 56\rangle}$}} = "f",
(0,0.4) *+{\scriptstyle{\langle 26 \rangle}} = "g",
(0.34641,-0.25) *+{\scriptstyle{\langle 24 \rangle}} = "h",
(-0.34641,-0.25) *+{\scriptstyle{\langle 46 \rangle}} = "i",
"h", {\ar"i"},
"g", {\ar"h"},
"i", {\ar"g"},
"g", {\ar"a"},
"b", {\ar"g"},
"h", {\ar"c"},
"d", {\ar"h"},
"i", {\ar"e"},
"f", {\ar"i"},
"e", {\ar@{.>}"f"},
"a", {\ar@{.>}"b"},
"c", {\ar@{.>}"d"}
\end{xy}
\\
(d) & (e) & (f)
\\
\\
\\
\begin{xy} 0;<53pt,0pt>:<0pt,53pt>::
(-0.5,\halfrootthree) *+{\framebox[5ex]{$\scriptstyle{\langle 3456\rangle}$}} = "a",
(0.5,\halfrootthree) *+{\framebox[5ex]{$\scriptstyle{\langle 1456\rangle}$}} = "b",
(1,0) *+{\framebox[5ex]{$\scriptstyle{\langle 1256\rangle}$}} = "c",
(0.5,-\halfrootthree) *+{\framebox[5ex]{$\scriptstyle{\langle 1236\rangle}$}} = "d",
(-0.5,-\halfrootthree) *+{\framebox[5ex]{$\scriptstyle{\langle 1234\rangle}$}} = "e",
(-1,0) *+{\framebox[5ex]{$\scriptstyle{\langle 2345\rangle}$}} = "f",
(0.25,0.4) *+{\scriptstyle{\langle 2456 \rangle}} = "g",
(0.1,-0.05) *+{\scriptstyle{\langle 2356 \rangle}} = "h",
(-0.35,-0.4) *+{\scriptstyle{\langle 2346 \rangle}} = "i",
"i", {\ar"h"},
"h", {\ar"g"},
"g", {\ar"a"},
"b", {\ar"g"},
"g", {\ar"c"},
"c", {\ar"h"},
"h", {\ar"d"},
"d", {\ar"i"},
"i", {\ar"e"},
"f", {\ar"i"},
"e", {\ar@{.>}"f"},
"a", {\ar@{.>}"b"}
\end{xy}
&
\begin{xy} 0;<53pt,0pt>:<0pt,53pt>::
(-0.5,\halfrootthree) *+{\framebox[5ex]{$\scriptstyle{\langle 3456\rangle}$}} = "a",
(0.5,\halfrootthree) *+{\framebox[5ex]{$\scriptstyle{\langle 1456\rangle}$}} = "b",
(1,0) *+{\framebox[5ex]{$\scriptstyle{\langle 1256\rangle}$}} = "c",
(0.5,-\halfrootthree) *+{\framebox[5ex]{$\scriptstyle{\langle 1236\rangle}$}} = "d",
(-0.5,-\halfrootthree) *+{\framebox[5ex]{$\scriptstyle{\langle 1234\rangle}$}} = "e",
(-1,0) *+{\framebox[5ex]{$\scriptstyle{\langle 2345\rangle}$}} = "f",
(0.2,0.4) *+{\scriptstyle{\langle 2456 \rangle}} = "g",
(0,0) *+{\scriptstyle{\langle 2356 \rangle}} = "h",
(-0.2,-0.4) *+{\scriptstyle{\langle 1235 \rangle}} = "i",
"h", {\ar"i"},
"h", {\ar"g"},
"g", {\ar"a"},
"b", {\ar"g"},
"g", {\ar"c"},
"c", {\ar"h"},
"i", {\ar"d"},
"f", {\ar"h"},
"e", {\ar"i"},
"i", {\ar"f"},
"d", {\ar@{.>}"e"},
"a", {\ar@{.>}"b"}
\end{xy}
&
\begin{xy} 0;<53pt,0pt>:<0pt,53pt>::
(-0.5,\halfrootthree) *+{\framebox[5ex]{$\scriptstyle{\langle 3456\rangle}$}} = "a",
(0.5,\halfrootthree) *+{\framebox[5ex]{$\scriptstyle{\langle 1456\rangle}$}} = "b",
(1,0) *+{\framebox[5ex]{$\scriptstyle{\langle 1256\rangle}$}} = "c",
(0.5,-\halfrootthree) *+{\framebox[5ex]{$\scriptstyle{\langle 1236\rangle}$}} = "d",
(-0.5,-\halfrootthree) *+{\framebox[5ex]{$\scriptstyle{\langle 1234\rangle}$}} = "e",
(-1,0) *+{\framebox[5ex]{$\scriptstyle{\langle 2345\rangle}$}} = "f",
(0,0.4) *+{\scriptstyle{\langle 2456 \rangle}} = "g",
(0.34641,-0.25) *+{\scriptstyle{\langle 1246 \rangle}} = "h",
(-0.34641,-0.25) *+{\scriptstyle{\langle 2346 \rangle}} = "i",
"h", {\ar"i"},
"g", {\ar"h"},
"i", {\ar"g"},
"g", {\ar"a"},
"b", {\ar"g"},
"h", {\ar"c"},
"d", {\ar"h"},
"i", {\ar"e"},
"f", {\ar"i"},
"e", {\ar@{.>}"f"},
"a", {\ar@{.>}"b"},
"c", {\ar@{.>}"d"}
\end{xy}
\\
(g) & (h) & (i)
\\
\end{tabular}
\caption{The three types of (reduced, perfectly orientable, bipartite) plabic
graphs corresponding to $km$-dimensional cells of $\Gr(k,n)_{\ge 0}$ for
$k=2$, $m=4$ and $n=6$ are shown in (a)--(c).
The associated input and output clusters (see text) are shown
in (d)--(f) and (g)--(i) respectively.  Lines connecting two frozen
nodes are usually omitted, but we include in (g)--(i) the dotted lines
(having ``black on the right'' in the dual plabic graph) that
encode~(\ref{eq:example1}),
(\ref{eq:example2}) and~(\ref{eq:example3}) (up to signs).}
\label{fig:n6k2}
\end{figure}

Modulo dihedral transformations on the six external points,
there are a total of four different types of plabic graph to consider.
We begin with the three graphs shown in Fig.~\ref{fig:n6k2} (a)--(c),
which have $k=2$.
These all correspond to the top cell of $\Gr(2,6)_{\ge 0}$ and
are related to each other by square moves.
Specifically, performing a square move on $f_2$ of graph (a)
yields  graph (b), while performing a square move
on $f_4$ of graph (a) yields graph (c).
This contrasts with more general cases, for example
those considered in the next two sections, where we are in general
interested in lower-dimensional cells.

The solution for the face weights of graph (a) (the same as~(\ref{eq:examplegraph}))
were already given in~(\ref{eq:example1}), and those of graphs (b) and (c) are derived
in~(\ref{eq:example2}) and~(\ref{eq:example3}) of Appendix~\ref{sec:details}.
The latter two can alternatively be derived from the former via
the square move rule (see~\cite{Postnikov,ArkaniHamed:2012nw}).  In particular, for graph (b) we have
\begin{equation}
\begin{aligned}
\begin{split}
f_0^{(b)} &= f_0^{(a)} (1 + f_2^{(a)}) \,, \\
f_1^{(b)} &= \frac{f_1^{(a)}}{1 + 1/f_2^{(a)}} \,,
\end{split}~~
\begin{split}
f_2^{(b)} &= \frac{1}{f_2^{(a)}} \,,
\end{split}~~
\begin{split}
f_3^{(b)} &= f_3^{(a)} (1 + f_2^{(a)}) \,, \\
f_4^{(b)} &= \frac{f_4^{(a)}}{1+1/f_2^{(a)}} \,,
\end{split}
\end{aligned}
\end{equation}
with $f_5$, $f_6$, $f_7$ and $f_8$ unchanged,
while for graph (c) we have
\begin{equation}
\begin{aligned}
\begin{split}
f_2^{(c)} &= f_2^{(a)} (1 + f_4^{(a)}) \,, \\
f_3^{(c)} &= \frac{f_3^{(a)}}{1 + 1/f_4^{(a)}} \,,
\end{split}~~
\begin{split}
f_4^{(c)} &= \frac{1}{f_4^{(a)}} \,,
\end{split}~~
\begin{split}
f_5^{(c)} &= f_5^{(a)} (1 + f_4^{(a)}) \,, \\
f_6^{(c)} &= \frac{f_6^{(a)}}{1+1/f_4^{(a)}} \,,
\end{split}
\end{aligned}
\end{equation}
with $f_0$, $f_1$, $f_7$ and $f_8$ unchanged.

To every plabic graph one can naturally associate a quiver
with nodes labeled by Pl\"ucker coordinates of $\Gr(k,n)$.  In
Fig.~\ref{fig:n6k2} (d)--(f)
we display these quivers for the graphs under consideration,
following the source-labeling convention
of~\cite{muller2016cluster,SSBW} (see also~\cite{Paulos:2014dja}).
Because in this case each graph corresponds to the top cell
of $\Gr(2,6)_{\ge 0}$, each labeled quiver is a seed of the
$\Gr(2,6)$ cluster algebra.  More generally we will have graphs
corresponding to lower-dimensional cells, whose labeled quivers
are seeds of subalgebras of $\Gr(k,n)$.

Henceforth we refer to a labeled quiver associated to a plabic
graph in this manner as an \emph{input cluster},
taking the point of view that solving the
equations $C \cdot Z = 0$
associates a collection of functions on $\Gr(m,n)$ to
every such input.
As we now explain,
for these three graphs there
is also a natural way to graphically
organize the structure of each of~(\ref{eq:example1}),
(\ref{eq:example2}) and~(\ref{eq:example3})
in terms of an \emph{output cluster}, although in the next
section we will see that this does not work for more general graphs.

First of all we note from
(\ref{eq:example2}) and~(\ref{eq:example3}) that, like
what happened for graph (a) considered in the previous
section, each face weight evaluates (up to sign) to a product
of powers of $\Gr(4,6)$ cluster variables.  Second,
again we see that for each graph the collection of
variables that appear precisely constitutes a single cluster
of $\Gr(4,6)$: suppressing in each case the six frozen variables,
we find $\ket{2346}$, $\ket{2356}$ and
$\ket{2456}$ for graph (a);
$\ket{1235}$, $\ket{2356}$ and $\ket{2456}$ for graph (b);
and $\ket{1456}$, $\ket{2346}$, and $\ket{2456}$ for graph (c).
Finally, in each case there is a unique way to label
the nodes of the quiver not with cluster variables of the
``input'' cluster algebra $\Gr(2,6)$, as we have
done in Fig.~\ref{fig:n6k2} (d)--(f), but with cluster
variables of the ``output'' cluster algebra $\Gr(4,6)$.
We show these output clusters
in Fig.~\ref{fig:n6k2} (g)--(i), using the
convention that the face weight (also known as the cluster
$\mathcal{X}$-variable) attached to node $i$
is $\prod_j a_j^{b_{ji}}$ where $b_{ji}$ is the (signed)
number of arrows from $j$ to $i$.

For the sake of completeness, we note that there is also (modulo
$\mathbb{Z}_6$ cyclic transformations) a single relevant graph with $k=1$:
\begin{center}
\includegraphics[width=1.5in]{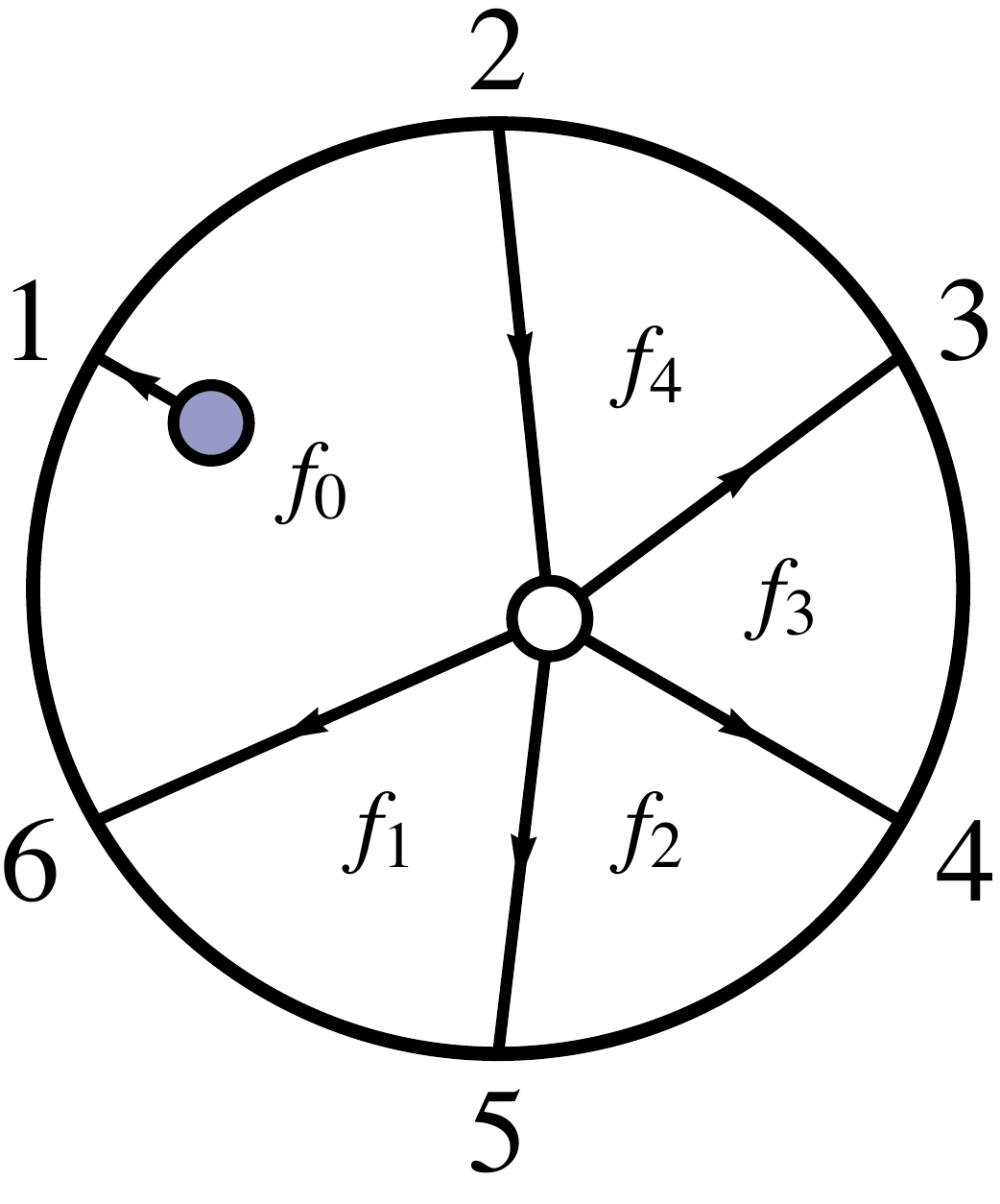}
\end{center}
for which the boundary measurement is
\begin{align}
C = \left(\begin{matrix}
0 & 1 & f_0 f_1 f_2 f_3 & f_0 f_1 f_2 & f_0 f_1 & f_0
\end{matrix}\right)
\end{align}
and the solution to $C \cdot Z = 0$ is given by
\begin{equation}
\begin{aligned}
f_0 &= \frac{\ket{2345}}{\ket{3456}}\,,&
f_1 &=-\frac{\ket{2346}}{\ket{2345}}\,,&
f_2 &=-\frac{\ket{2356}}{\ket{2346}}\,,&
f_3 &=-\frac{\ket{2456}}{\ket{2356}}\,,&
f_4 &=-\frac{\ket{3456}}{\ket{2456}}\,.
\end{aligned}
\end{equation}
Again the face weights evaluate (up to signs) to simple ratios of $\Gr(4,6)$ cluster variables,
though in this case both the input and output quivers are trivial.
This graph is
an example of the general feature that one can always uplift an $n$-point plabic
graph relevant to our analysis to any value of $n' > n$ by inserting any number of
black lollipops.  (Graphs with white lollipops never admit solutions to
$C \cdot Z = 0$ for generic $Z$.)  In the language of symbology, this is in accord with the
expectation that the symbol alphabet of an $n'$-particle amplitude always contains the
$\mathbb{Z}_{n'}$ cyclic closure of the symbol alphabet of the corresponding $n$-particle
amplitude.

In this section we have seen that solving $C \cdot Z = 0$ induces a map from
clusters of $\Gr(2,6)$ (or subalgebras thereof) to clusters of $\Gr(4,6)$ (or subalgebras thereof).
Of course, these two algebras are in any case naturally isomorphic.
Although we leave a more detailed exposition for future work,
we have also checked for $m=2$ and $n \le 10$ that every appropriate plabic
graph of $\Gr(k,n)$ maps to a cluster of $\Gr(2,n)$ (or a subalgebra thereof), and
moreover that this map is onto (every cluster of $\Gr(2,n)$
is obtainable from some plabic graph of $\Gr(k,n)$).
However for $m > 2$ the situation is more complicated, as we see in the next section.

\section{Towards Non-Cluster Variables}
\label{sec:noncluster}

Here we discuss some features of
graphs for which the solution to $C \cdot Z = 0$ involves
quantities that are not cluster variables of $\Gr(m,n)$.
A simple example for $k=2$, $m=3$, $n=6$
is the graph
\begin{equation}
\begin{aligned}
\includegraphics[width=1.5in]{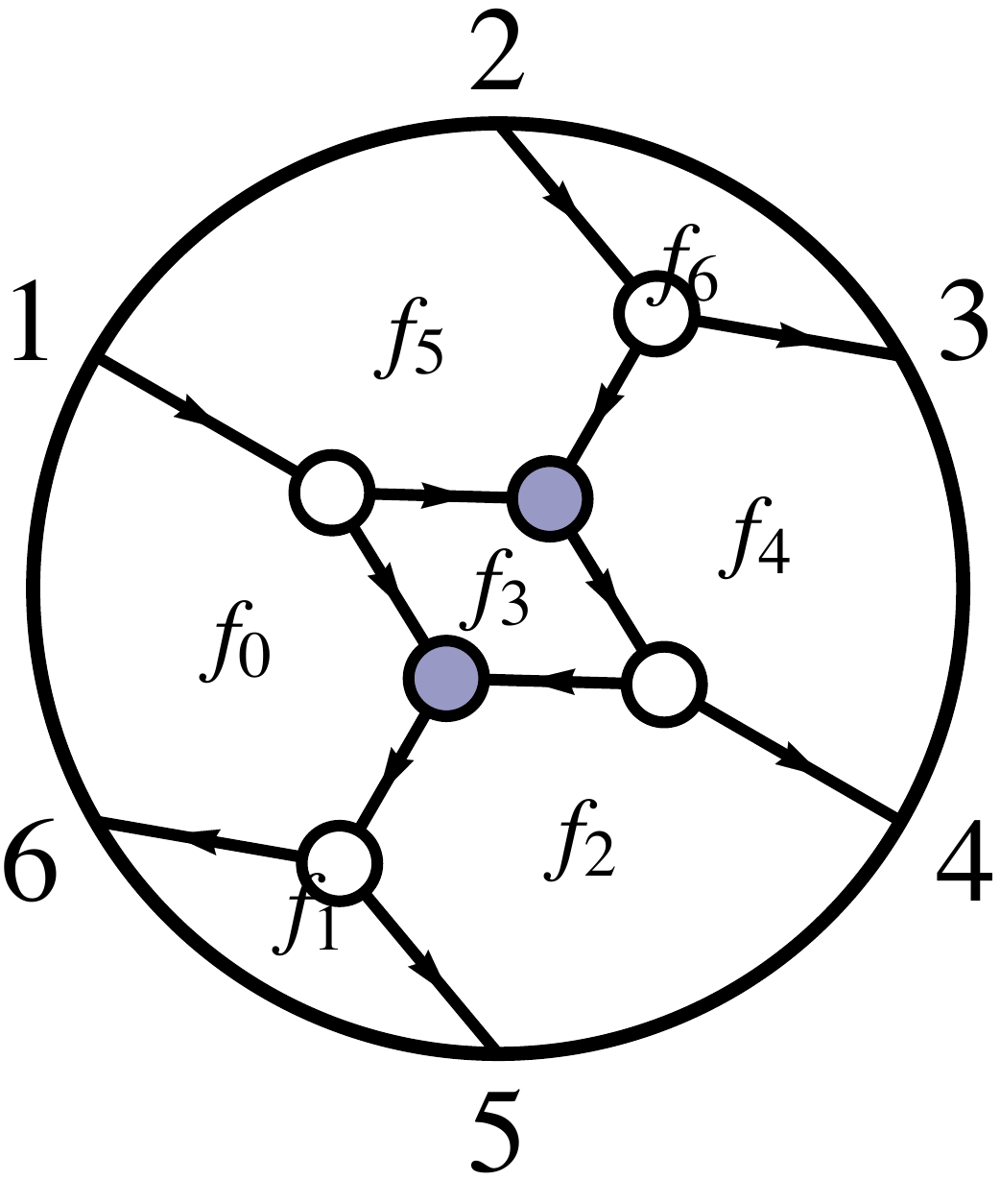}
\end{aligned}
\label{eq:towardsexample}
\end{equation}
whose boundary measurement has the form~(\ref{eq:cmatrixform})
with
\begin{equation}
\begin{aligned}
c_{13} &= - 0 \,, &
c_{15} &= - f_0 f_1 (1 + f_3)\,, &
c_{23} &= f_0 f_1 f_2 f_3 f_4 f_5\,, &
c_{25} &= f_0 f_1 f_3 f_5\,,  \\
c_{14} &= - f_0 f_1 f_2 f_3\,, &
c_{16} &= - f_0 (1 + f_3)\,, &
c_{24} &= f_0 f_1 f_2 f_3 f_5\,,  &
c_{26} &= f_0 f_3 f_5\,.
\end{aligned}
\end{equation}
The solution to $C \cdot Z = 0$ is given by
\begin{equation}
\begin{aligned}
f_0 &= \frac{\ket{123} \ket{145}}{\ket{1 \times 4, 2 \times 3, 5 \times 6}} \,, &
f_1 &=-\frac{\ket{146}}{\ket{145}} \,, \\
f_2 &= \frac{\ket{1 \times 4, 2 \times 3, 5 \times 6}}{\ket{234} \ket{146}} \,, &
f_3 &=-\frac{\ket{234} \ket{156}}{\ket{123}\ket{456}} \,, \\
f_4 &=-\frac{\ket{124} \ket{456}}{\ket{1 \times 4, 2 \times 3, 5 \times 6}} \,, &
f_5 &= \frac{\ket{1 \times 4, 2 \times 3, 5 \times 6}}{\ket{134}\ket{156}} \,, \\
f_6 &=-\frac{\ket{134}}{\ket{124}}\,,
\end{aligned}
\label{eq:badsolution}
\end{equation}
which involves four mutable cluster variables
$\ket{124}$, $\ket{134}$, $\ket{145}$, and $\ket{146}$
which comprise a cluster of the $\Gr(3,6)$ algebra,
together with the quantity
\begin{align}
\ket{1 \times 4, 2 \times 3, 5 \times 6} = \ket{123} \ket{456} - \ket{234} \ket{156}\,,
\label{eq:offender}
\end{align}
which is not a cluster variable of $\Gr(3,6)$.

We can gain some insight into the origin of~(\ref{eq:offender})
by considering what happens under a square move on $f_3$.
This transforms the face weights to
\begin{equation}
\begin{aligned}
f_0 &= \frac{\ket{145}}{\ket{456}} \,, &
f_1 &=-\frac{\ket{146}}{\ket{145}} \,, &
f_2 &=-\frac{\ket{156}}{\ket{146}} \,, &
f_3 &=-\frac{\ket{123} \ket{456}}{\ket{234} \ket{156}} \,, \\
f_4 &=-\frac{\ket{124}}{\ket{123}} \,, &
f_5 &=-\frac{\ket{234}}{\ket{134}} \,, &
f_6 &=-\frac{\ket{134}}{\ket{124}}\,,
\end{aligned}
\label{eq:badoutput}
\end{equation}
leaving four mutable cluster variables $\ket{124}$, $\ket{134}$, $\ket{145}$, and $\ket{146}$ which comprise a cluster
of the $\Gr(3,6)$ algebra.
However, it is not possible to associate a labeled ``output'' quiver to~(\ref{eq:badoutput}) in the usual way,
as we did for the examples in the previous section.

To turn this story around:
had we started not with~(\ref{eq:towardsexample}) but with its
square-moved partner, we would have encountered~(\ref{eq:badoutput})
and then noted that performing a square
move back to~(\ref{eq:towardsexample}) would necessarily introduce the multiplicative factor
\begin{align}
1 + f_3 = - \frac{\ket{1 \times 4, 2 \times 3, 5 \times 6}}{\ket{234}\ket{156}}
\end{align}
into four of the face weights.

The example considered in this section provides an opportunity to comment
on the connection of our work to the study of cluster adjacency
for Yangian invariants.  In~\cite{Drummond:2018dfd,Mago:2019waa} it was noted
in several examples, and conjectured to be true in general,
that the set of factors appearing in the denominator of any Yangian invariant
with intersection number 1
are cluster variables of $\Gr(4,n)$ that appear together in a cluster.
This was proven to be true for all Yangian invariants in the $m=2$
toy model of SYM theory in~\cite{Lukowski:2019sxw} and for all
$m=4$
N${}^2$MHV Yangian invariants in~\cite{Gurdogan:2020tip}.
We recall from~\cite{ArkaniHamed:2012nw,Arkani-Hamed:2017vfh}
that the Yangian
invariant associated to a plabic graph (or, to use essentially
equivalent language, the form associated to an on-shell diagram)
is given by $d \log f_1 \wedge \cdots \wedge d \log f_d$.
One of our motivations for studying the conjecture that
the face weights associated to any plabic graph
always evaluate on the solution of $C \cdot Z = 0$ to products
of powers of cluster variables was that it would immediately
imply cluster adjacency for Yangian invariants.
Although
the graph~(\ref{eq:towardsexample})
violates this stronger
conjecture, it does not violate cluster adjacency
because on-shell forms are invariant
under square moves~\cite{ArkaniHamed:2012nw}.
Therefore, even though $\ket{1 \times 4, 2 \times 3, 5 \times 6}$
appears in individual face weights of~(\ref{eq:badsolution}), it must
drop out of the associated on-shell form because it is absent from~(\ref{eq:badoutput}).

\section{Algebraic Eight-Particle Symbol Letters}
\label{sec:rootsection}

One reason it is obvious that the solutions of $C \cdot Z = 0$ cannot always
be written in terms of cluster variables of $\Gr(m,n)$ is that
for graphs with intersection number greater than 1 the solutions will
necessarily involve algebraic functions of Pl\"ucker coordinates, whereas
cluster variables are always rational.

The simplest example of this phenomenon occurs for $k=2$, $m=4$ and $n=8$,
for which there are four relevant plabic graphs in two cyclic classes.
Let us start with
\begin{equation}
\begin{aligned}
\includegraphics[width=1.5in]{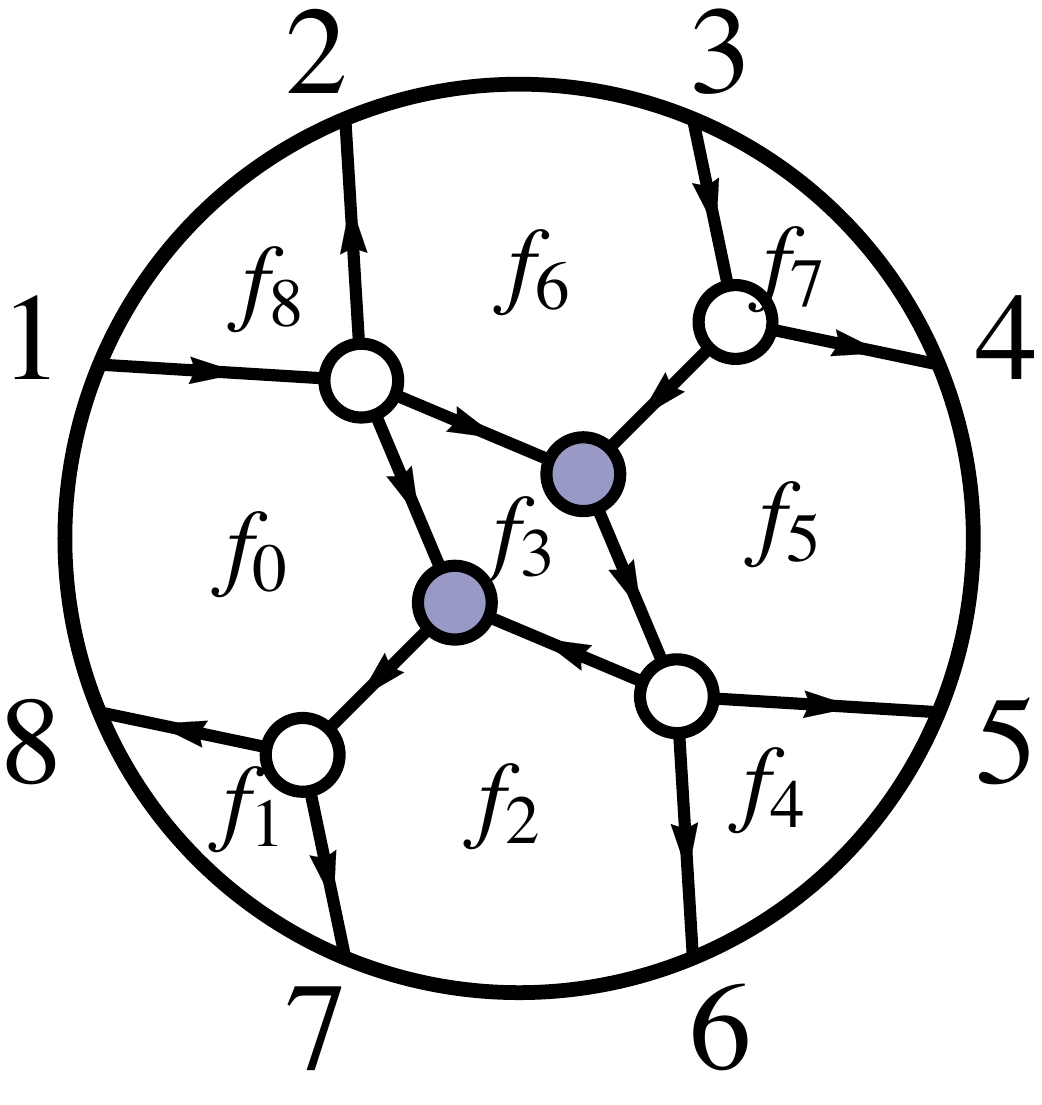}
\end{aligned}
\label{eq:sqrtplabic}
\end{equation}
which has boundary measurement
\begin{align}
C = \left(\begin{matrix}
1 & c_{12} & 0 &  c_{14} &  c_{15} &  c_{16} &  c_{17} &  c_{18} \\
0 & c_{32} & 1 & c_{34} & c_{35} & c_{36} & c_{37} & c_{38}
\end{matrix}\right)
\end{align}
with
\begin{align}
c_{12} &= f_0 f_1 f_2 f_3 f_4 f_5 f_6 f_7 \,, &
c_{14} &= -0 \,, &
c_{15} &= - f_0 f_1 f_2 f_3 f_4\,,  \\
c_{16} &= -f_0 f_1 f_2 f_3\,, &
c_{17} &= -f_0 f_1 (1+f_3)\,, &
c_{18} &= -f_0 (1+f_3)\,, \\
c_{32} &= 0\,, &
c_{34} &= f_0 f_1 f_2 f_3 f_4 f_5 f_6 f_8\,, &
c_{35} &= f_0 f_1 f_2 f_3 f_4 f_6 f_8\,, \\
c_{36} &= f_0 f_1 f_2 f_3 f_6 f_8\,, &
c_{37} &= f_0 f_1 f_3 f_6 f_8\,, &
c_{38} &= f_0 f_3 f_6 f_8\,.
\end{align}
The solution to $C \cdot Z = 0$ for generic $Z \in \Gr(4,8)$ can be written as
\begin{equation}
\begin{aligned}
f_0 &= \sqrt{\frac{\ket{7(12)(34)(56)}\,\ket{1234}}{a_5\, \ket{2(34)(56)(78)}\,\ket{3478}}} \,, &
f_5 &= \sqrt{\frac{a_1 a_6 a_9\, \ket{3(12)(56)(78)}\,\ket{5678}}{a_4 a_7\, \ket{6(12)(34)(78)}\,\ket{3478}}}\,, \\
f_1 &=-\sqrt{\frac{a_7\, \ket{8(12)(34)(56)}}{\ket{7(12)(34)(56)}} }\,, &
f_6 &=-\sqrt{\frac{a_3\, \ket{1(34)(56)(78)}\,\ket{3478}}{a_2\, \ket{4(12)(56)(78)}\,\ket{1278}}}\,, \\
f_2 &=-\sqrt{\frac{a_4\, \ket{5(12)(34)(78)}\, \ket{3478}}{a_8\, \ket{8(12)(34)(56)}\,\ket{3456}}}\,, &
f_7 &=-\sqrt{\frac{a_2\, \ket{4(12)(56)(78)}}{a_1 \ket{3(12)(56)(78)}}}\,, \\
f_3 &= \sqrt{\frac{a_8\, \ket{1278}\, \ket{3456}}{a_9\, \ket{1234}\, \ket{5678}}}\,, &
f_8 &=-\sqrt{\frac{a_5\, \ket{2(34)(56)(78)}}{a_3\, \ket{1(34)(56)(78)}}}\,, \\
f_4 &=-\sqrt{\frac{\ket{6(12)(34)(78)}}{a_6\, \ket{5(12)(34)(78)}}}\,,
\end{aligned}
\label{eq:last}
\end{equation}
where
\begin{align}
\langle a(bc)(de)(fg)\rangle \equiv
\ket{abde}\ket{acfg} - \ket{abfg}\ket{acde}
\label{eq:bracketdef}
\end{align}
and the nine $a_i$ provide a (multiplicative) basis for
the algebraic letters of the eight-particle symbol alphabet
that contain the four-mass box square root $\sqrt{\Delta_{1357}}$,
as reviewed in Appendix~\ref{sec:roots}.

The nine face weights shown in~(\ref{eq:last}) satisfy
$\prod f_\alpha = 1$ so only eight are multiplicatively independent.
It is easy to check that they remain multiplicatively independent
if one sets all of the Pl\"ucker coordinates and brackets
of the form~(\ref{eq:bracketdef}) to one.
This means that the projection of this eight-dimensional space onto
the nine-dimensional space spanned by the nine algebraic letters
is eight dimensional (and not smaller).
We could try building an eight-particle alphabet by taking
any subset of eight of the face weights as basis elements (i.e., letters),
but we would always be one letter short of spanning the full algebraic
space.

Fortunately there is a second plabic graph relevant
to $\sqrt{\Delta_{1357}}$: the one obtained by performing
a square move on $f_3$ of~(\ref{eq:sqrtplabic}).  As is by now
familiar, performing the square move introduces one
new multiplicative factor into the face weights:
\begin{align}
1 + f_3 = \sqrt{\frac{\ket{1256} \ket{3478}}{a_9 \ket{1234} \ket{5678}}}\,,
\end{align}
which precisely supplies the ninth, missing letter!  To summarize:
the union of the nine face weights associated to the graph~(\ref{eq:sqrtplabic}), and the nine associated to its square-move partner, multiplicatively
span the nine-dimensional space of $\sqrt{\Delta_{1357}}$-containing symbol letters
in the eight-particle alphabet of~\cite{Zhang:2019vnm}.

The same story applies to the graphs obtained by cycling the external
indices on~(\ref{eq:sqrtplabic}) by one---their face weights provide all
nine algebraic letters involving $\sqrt{\Delta_{2468}}$.

Of course it would be very interesting to thoroughly study
the numerous plabic graphs relevant to $m=4$, $n=8$ that
have intersection number 1.  In particular it would be interesting
to see if they encode all 180 of the rational
(i.e., $\Gr(4,8)$ cluster variable)
symbol letters of~\cite{Zhang:2019vnm}, and
whether they generate additional cluster variables
such as those obtained from the constructions of~\cite{Drummond:2019cxm,Arkani-Hamed:2019rds,Henke:2019hve}.

Before concluding this section let us comment briefly on ``$k$'' since one
may be confused why the plabic graph~(\ref{eq:sqrtplabic}), which has $k=2$
and is therefore associated to an N${}^2$MHV leading singularity,
could be relevant for symbol alphabets of NMHV amplitudes.
The symbol letters of an N${}^k$MHV amplitude reveal
all of its singularities, including multiple
discontinuities that can be accessed only after a suitable analytic
continuation.  Physically these are computed by cuts involving
lower-loop amplitudes that can have $k'>k$. Indeed
the expectation that symbol letters of lower-loop higher-$k$ amplitudes
influence those of higher-loop lower-$k$ amplitudes
is manifest in the $\overline{Q}$-bar equation
technology~\cite{CaronHuot:2011ky,CaronHuot:2011kk,Bullimore:2011kg}
underlying the computation of~\cite{Zhang:2019vnm}.
Moreover there is indirect evidence~\cite{Prlina:2018ukf}
that the symbol alphabet of the $L$-loop $n$-particle N${}^k$MHV amplitude in
SYM theory is independent of both $k$ and $L$ (beyond certain
accidental shortenings that may occur for small $k$ or $L$).
This suggests that for the purpose of applying our construction
to ``\emph{the} $n$-particle symbol alphabet''
one should
consider all relevant $n$-point plabic graphs regardless of $k$.

\section{Discussion}

The problem of ``explaining'' the symbol alphabets of $n$-particle
amplitudes in SYM theory apparently
requires, for $n>7$, a mechanism for identifying
finite sets of functions on $\Gr(4,n)$ that include some subset of the
cluster variables of the associated cluster algebra, together with
certain non-cluster variables that are algebraic functions of
the Pl\"ucker coordinates.

In this paper we have initiated the study of one candidate mechanism
that manifestly satisfies both criteria and may be of independent
mathematical interest.
Specifically, to every (reduced, perfectly oriented) plabic
graph of $\Gr(k,n)_{\ge 0}$
that parameterizes a cell of dimension $mk$, one can naturally associate
a collection of $mk$ functions of Pl\"ucker coordinates on $\Gr(m,n)$.

We have seen that for some graphs the output of this procedure is
naturally associated to a seed of the $\Gr(m,n)$ cluster algebra;
for some graphs
the output is a cluster's worth of cluster variables that
do not correspond to a seed but rather behave ``badly'' under
mutations (this means they transform into things
which are not cluster variables under square moves on the
input plabic graph);
and finally for some graphs the output involves non-cluster variables
including, when the intersection number is greater than 1,
algebraic functions.

We leave a more thorough investigation of this problem for
future work.  The ``smoking gun'' that this procedure may be relevant
to symbol alphabets in SYM theory is provided by the example discussed
in Sec.~\ref{sec:rootsection}, which successfully postdicts precisely
the 18 multiplicatively independent algebraic letters that were recently
found to appear in the two-loop eight-particle
NMHV amplitude~\cite{Zhang:2019vnm}.
Our construction provides an alternative to the
similar postdiction made recently in~\cite{Drummond:2019cxm}.

It is interesting to note that
since for $m=4$, $n=8$
there are no other relevant
plabic graphs having intersection number $> 1$, beyond
those already considered Sec.~\ref{sec:rootsection}, our construction
has no room for any additional algebraic letters for eight-particle amplitudes.
Therefore, if it is true that the face weights of plabic
graphs, evaluated on the locus $C \cdot Z = 0$, provide symbol
alphabets for general amplitudes, then it necessarily follows
that no eight-particle amplitude, at any loop order, can have
any algebraic symbol letters beyond the 18 discovered in~\cite{Zhang:2019vnm}.

At first glance this rigidity seems to stand in contrast
to the constructions of~\cite{Arkani-Hamed:2019rds,Drummond:2019cxm,Henke:2019hve}
which each involve some amount of choice---having
to do with how coarse or fine one chooses one's tropical fan, or
equivalently how many factors to include in the Minkowski sum
when building the dual polytope.
But in fact our construction has a choice with a similar smell.
When we say that we start with the $C$-matrix associated to a plabic
graph, that automatically restricts us to very
special clusters of $\Gr(k,n)$---those that contain
only Pl\"ucker coordinates.
Clusters containing more complicated,
non-Pl\"ucker cluster variables are not associated to plabic graphs.
One certainly could contemplate solving the $C \cdot Z = 0$ equations
for $C$ given by a ``non-plabic'' cluster parameterization of some
cell of $\Gr(k,n)_{\ge 0}$ and it would be interesting to map
out the landscape of possibilities.

It has been a long-standing problem to understand
the precise connection between the $\Gr(k,n)$ cluster structure exhibited~\cite{ArkaniHamed:2012nw}
at the level of integrands in SYM theory and the
$\Gr(4,n)$ cluster structure exhibited~\cite{Golden:2013xva} by integrated amplitudes.
It was pointed out in~\cite{NimaTalk} that the $C \cdot Z = 0$ equations provide
a concrete link between the two, and our results
shed some initial light on this
intriguing but still very mysterious problem.
Since plabic graphs categorize the singularities of
integrands of SYM theory, whereas symbol letters characterize
the singularities of integrated amplitudes,
perhaps one can think of the ``input'' and ``output'' clusters described
in Sec.~\ref{sec:sixparticle} as ``integrand'' and ``integrated'' clusters.
(See the last paragraph of Sec.~\ref{sec:rootsection} for some comments
on why $k$ ``disappears'' upon integration.)
Although we have seen that the latter are not in general clusters
at all, the example of Sec.~\ref{sec:rootsection} suggests that they may be
even better:
exactly what is needed for the symbol alphabets of SYM theory.

\bigskip

\noindent
{\bf Note Added:}  The preprint~\cite{HL} appeared on arXiv shortly
after and
has significant overlap with the result presented in this note.

\acknowledgments

We are grateful to N.~Arkani-Hamed for numerous encouraging discussions
and to J.~Bourjaily for helpful correspondence.
This work was supported in part by the US Department of Energy under contract {DE}-{SC}0010010 Task A and by Simons Investigator Award \#376208 (AV). Plabic graphs were drawn with the help of~\cite{Bourjaily:2012gy}.

\appendix

\section{Some Six-Particle Details}
\label{sec:details}

Here we assemble some details of the calculation for graphs (b) and (c) of Fig.~\ref{fig:n6k2}.
The boundary measurement for graph (b) has the form~(\ref{eq:cmatrixform}) with
\begin{equation}
\begin{aligned}
c_{13} &= - f_0 f_1 f_2 f_3 f_4 f_5 f_6 \,, &
c_{23} &= f_0 f_1 f_2 f_3 f_4 f_5 f_6 f_8\,, \\
c_{14} &=- f_0 f_1 f_2 f_3 f_4 (1 + f_6)\,, &
c_{24} &= f_0 f_1 f_2 f_3 f_4 f_6 f_8\,, \\
c_{15} &=- f_0 f_1 (1 + f_4 + f_2 f_4 + f_4 f_6 + f_2 f_4 f_6) \,, &
c_{25} &= f_0 f_1 f_4 f_6 f_8 ( 1 + f_2) \,, \\
c_{16} &=-f_0 (1 + f_4 + f_4 f_6)\,, &
c_{26} &= f_0 f_4 f_6 f_8 \,,
\end{aligned}
\end{equation}
and the solution to $C \cdot Z = 0$ is given by
\begin{equation}
\begin{aligned}
f^{(b)}_0 &= - \frac{\ket{1235}}{\ket{2356}}\,, &
f^{(b)}_1 &= - \frac{\ket{1236}}{\ket{1235}}\,, &
f^{(b)}_2 &= \frac{\ket{1234}\ket{2356}}{\ket{2345}\ket{1236}}\,, \\
f^{(b)}_3 &= - \frac{\ket{1235}}{\ket{1234}}\,, &
f^{(b)}_4 &= \frac{\ket{2345}\ket{1256}}{\ket{1235}\ket{2456}}\,, &
f^{(b)}_5 &=-  \frac{\ket{2456}}{\ket{2356}}\,, \\
f^{(b)}_6 &= \frac{\ket{2356}\ket{1456}}{\ket{3456}\ket{1256}}\,. &
f^{(b)}_7 &= - \frac{\ket{3456}}{\ket{2456}}\,, &
f^{(b)}_8 &= - \frac{\ket{2456}}{\ket{1456}}\,.
\end{aligned}
\label{eq:example2}
\end{equation}
The boundary measurement for graph (c) has
\begin{equation}
\begin{aligned}
c_{13} &= -f_0 f_1 f_2 f_3 f_4 f_5 f_6\,, &
c_{23} &= f_0 f_1 f_2 f_3 f_4 f_5 f_6 f_8\,, \\
c_{14} &= -f_0 f_1 f_2 f_3 (1 + f_6 + f_4 f_6)\,, &
c_{24} &= f_0 f_1 f_2 f_3 f_6 f_8 (1 + f_4)\,, \\
c_{15} &= -f_0 f_1 f_2 (1 + f_6)\,, &
c_{25} &= f_0 f_1 f_2 f_6 f_8\,, \\
c_{16} &= -f_0 (1 + f_2 + f_2 f_6)\,, &
c_{26} &= f_0 f_2 f_6 f_8\,,
\end{aligned}
\end{equation}
and the solution to $C \cdot Z = 0$ is
\begin{equation}
\begin{aligned}
f^{(c)}_0 &= - \frac{\ket{1234}}{\ket{2346}}\,, &
f^{(c)}_1 &= - \frac{\ket{2346}}{\ket{2345}}\,, &
f^{(c)}_2 &= \frac{\ket{2345}\ket{1246}}{\ket{1234}\ket{2456}}\,, \\
f^{(c)}_3 &= - \frac{\ket{1256}}{\ket{1246}}\,, &
f^{(c)}_4 &= \frac{\ket{2456}\ket{1236}}{\ket{2346}\ket{1256}}\,, &
f^{(c)}_5 &=-  \frac{\ket{1246}}{\ket{1236}}\,, \\
f^{(c)}_6 &= \frac{\ket{1456}\ket{2346}}{\ket{3456}\ket{1246}}\,, &
f^{(c)}_7 &= - \frac{\ket{3456}}{\ket{2456}}\,, &
f^{(c)}_8 &= - \frac{\ket{2456}}{\ket{1456}}\,.
\end{aligned}
\label{eq:example3}
\end{equation}

\section{Notation for Algebraic Eight-Particle Symbol Letters}
\label{sec:roots}

Here we review some details from~\cite{Zhang:2019vnm}
to set the notation used in Sec.~\ref{sec:rootsection}.
There are two basic square roots of four-mass box type that
appear in symbol letters of eight-particle amplitudes.  These are
$\sqrt{\Delta_{1357}}$ and $\sqrt{\Delta_{2468}}$ with
\begin{align}
\Delta_{1357} = (\ket{1256} \ket{3478} - \ket{1278} \ket{3456} -
\ket{1234} \ket{5678})^2 - 4 \ket{1234} \ket{3456} \ket{5678}
\ket{1278}
\end{align}
and $\Delta_{2468}$ given by cycling every index by 1 (mod 8).

The eight-particle symbol alphabet can be written in terms of
180 $\Gr(4,8)$ cluster variables, plus 9 letters that are rational
functions of Pl\"ucker coordinates and $\sqrt{\Delta_{1357}}$ and
another 9 that are rational
functions of Pl\"ucker coordinates and $\sqrt{\Delta_{2468}}$.
We focus on the first 9 as the latter is a cyclic copy of the same
story.

There are many different ways to write a basis for the eight-particle symbol alphabet
as the various letters one can form satisfy numerous multiplicative
identities among each other.  For the sake of definiteness we use
the basis provided in the
ancillary Mathematica file attached to~\cite{Zhang:2019vnm}.
The choice of basis made there starts by defining
\begin{equation}
\begin{aligned}
z &= \frac{1}{2} (1 + u - v + \sqrt{(1-u-v)^2-4 u v})\,,\\
\bar{z} &= \frac{1}{2} (1 + u - v - \sqrt{(1-u-v)^2-4 u v})
\end{aligned}
\label{eq:zs}
\end{equation}
in terms of
the familiar eight-particle cross ratios
\begin{align}
u = \frac{\ket{1278} \ket{3456}}{\ket{1256} \ket{3478}} \,,
\qquad
v = \frac{\ket{1234} \ket{5678}}{\ket{1256}\ket{3478}}\,.
\end{align}
Note that the square root appearing in~(\ref{eq:zs}) is
\begin{align}
\sqrt{(1 - u - v)^2 - 4 u v} = \frac{\sqrt{\Delta_{1357}}}{\ket{1256} \ket{3478}}\,.
\end{align}
Then a basis for the algebraic letters of the symbol alphabet
is given by
\begin{equation}
\begin{aligned}
a_1 &= \left.\frac{x_a - z}{x_a - \bar{z}}\right|_{i \to i + 6}\,,&
a_2 &= \left.\frac{x_b - z}{x_b - \bar{z}}\right|_{i \to i + 6}\,,&
a_3 &=-\left.\frac{x_c - z}{x_c - \bar{z}}\right|_{i \to i + 6}\,,\\
a_4 &=-\left.\frac{x_d - z}{x_d - \bar{z}}\right|_{i \to i + 4}\,,&
a_5 &=-\left.\frac{x_d - z}{x_d - \bar{z}}\right|_{i \to i + 6}\,,&
a_6 &= \left.\frac{x_e - z}{x_e - \bar{z}}\right|_{i \to i + 4}\,,\\
a_7 &= \left.\frac{x_e - z}{x_e - \bar{z}}\right|_{i \to i + 6}\,,&
a_8 &= \frac{z}{\bar{z}}\,,&
a_9 &= \frac{1-z}{1-\bar{z}}\,,\\
\end{aligned}
\end{equation}
where the $x$'s are defined in~(13) of~\cite{Zhang:2019vnm}.
While the overall sign of a symbol letter is irrelevant,
we have taken the liberty of putting a minus sign in front of $a_3$, $a_4$
and $a_5$ to ensure that each of the nine $a_i$, indeed each individual
factor appearing in~(\ref{eq:last}), is positive-valued for $Z \in \Gr(4,8)_{>0}$.

\end{document}